# Electrical control of metallic heavy-metal/ferromagnet interfacial states

Chong Bi[1*], Congli Sun[2], Meng Xu[1], Ty Newhouse-Illige[1], Paul M. Voyles[2], and Weigang Wang[1†]

[1]Department of Physics, University of Arizona, Tucson, Arizona 85721, USA.

[2]Department of Materials Science and Engineering, University of Wisconsin-Madison, Madison, WI 53706, USA

[*] cbi@email.arizona.edu

[†]wgwang@physics.arizona.edu

Voltage control effects provide an energy-efficient means of tailoring material properties, especially in highly integrated nanoscale devices. However, only insulating and semiconducting systems can be controlled so far. In metallic systems, there is no electric field due to electron screening effects and thus no such control effect exists. Here we demonstrate that metallic systems can also be controlled electrically through ionic not electronic effects. In a Pt/Co structure, the control of the metallic Pt/Co interface can lead to unprecedented control effects on the magnetic properties of the entire structure. Consequently, the magnetization and perpendicular magnetic anisotropy of the Co layer can be independently manipulated to any desired state, the efficient spin toques can be enhanced about 3.5 times, and the switching current can be reduced about one order of magnitude. This ability to control a metallic system may be extended to control other physical phenomena.

# I. Introduction

The heavy-metal (HM)/ferromagnet (FM)/oxide structure is one of the most important building blocks of spintronics, where both the HM/FM and FM/oxide interfaces result in various emergent magnetic phenomena, such as tunnel magnetoresistance (TMR) [1–4], perpendicular magnetic anisotropy (PMA) [1–8], Rashba spin-orbit coupling (RSOC) [9–12], and the Dzyaloshinskii-Moriya interaction (DMI) [13,14]. In particular, PMA is of great importance for future generations of high-density memory and logic applications. It has been demonstrated that PMA can be controlled by an electric field (EF) applied to the oxide layer [1,3,5,15,16], thereby opening a potential way to achieve ultra-low energy magnetization switching with very little Joule heating. Although both theoretical [6,17] and experimental [2,7,8] studies indicate that both interfaces contribute to PMA simultaneously, all EF-control studies only focus on control of the FM/oxide interface where EF can penetrate. The HM/FM interface at which no EF exists has not been explored.

There are two types of EF-control effects. One is the voltage controlled magnetic anisotropy (VCMA) effect with an electronic origin, where PMA can be modified by the EF-induced redistribution of electron densities among different $d$ orbitals in FMs [5,18–20], or through an EF-modified RSOC/DMI [9,10]. In VCMA, the magnetic anisotropy field of the FM layer can be effectively modified by an external EF [1,5,18,19,21,22], which has been successfully employed to achieve efficient switching in magnetic tunnel junctions [1,3], modulation of domain wall propagation [22], and control of the Curie temperature [21] and skyrmion bubbles [23]. The second type is of ionic origin, where the ionic motion of $O^{2-}$ driven by an EF is exploited to control the interfacial oxidation states of a FM [15,16,24–26]. Through this solid state electrochemical reaction at the FM/oxide interface, the saturation magnetization ($M_s$) of the FM can also be controlled. The ionic effect provides a much larger control over PMA, on the order of 10 pJ/Vm, as opposed to a few hundreds of fJ/Vm with VCMA [15,16]. Moreover, unlike electrons, ions can diffuse into a metallic system even without EFs, and thus the ionic effect may be employed to control the metallic HM/FM interface. It is known that, in addition to PMA, the HM/FM interface also determines spin-orbit torque (SOT) driven magnetic dynamics [27,28] and DMI effects [13,14] in these structures. Therefore, it would be highly advantageous if the HM/FM interface could also be controlled electrically, which will pave a new path to control PMA and simultaneously modulate the interfacial phenomena to improve the efficiency of SOT switching. However, the voltage controlled HM/FM interface has yet to be demonstrated.

Here we demonstrate voltage control of the HM/FM interface in a Pt/Co/GdO$_x$ structure for the first time. By combining independent control of the Co/GdO$_x$ interface, we have reached an unprecedentedly effective manipulation of PMA in HM/FM/oxide structures, greatly expanding the controllable magnetic states to those that cannot be achieved by controlling only the FM/oxide interface in previous studies. Moreover, both the PMA field ($H_k$) and $M_s$ can be controlled independently and the magnetism can be configured to any possible state, in sharp contrast to previous studies where $M_s$ and $H_k$ were always controlled together when only the FM/oxide interface was modulated [15,16]. The efficiency of SOTs is also controlled by voltage and the critical SOT switching current ($I_c$) is reduced by about one order of magnitude. These results highlight the unexplored voltage control of metallic HM/FM interfaces in electrical control of magnetism, which may be extended to control other interfacial effects [29–31].

## II. Methods and materials

*Sample fabrication.* Three batches of Pt/Co/GdO$_x$ samples with structures of Pt (10 nm)/Co (4 nm)/GdO$_x$ (50 nm), Pt (3 nm)/Co (0.9 nm)/GdO$_x$ (80 nm), and Pt (3 nm)/Co (0.6 nm)/GdO$_x$ (80 nm) were deposited on silicon substrates with a 300 nm thermally oxidized SiO$_2$ layer by magnetron sputtering at a base pressure of $2 \times 10^{-8}$ torr. The Pt (10 nm)/Co (4 nm)/GdO$_x$ (50 nm) samples with thick Pt and Co layers were used for observing EF-driven oxygen migration under Cross-sectional Scanning Transmission Electron Microscopy (STEM). The Pt (3 nm)/Co (0.9 nm)/GdO$_x$ (80 nm) and Pt (3 nm)/Co (0.6 nm)/GdO$_x$ (80 nm) samples that show PMA in the as-deposited state with much thinner Co layers were used for investigating EF-control of PMA and SOT switching, respectively. All metallic layers were dc sputtered and GdO$_x$ was reactively sputtered. The deposited continuous multilayers were then patterned into Hall bar structures. The residual areas are protected by 100 nm SiO$_2$. The top gate electrodes are made of Ta (5 nm)/Ru (100 nm).

*Electrical measurements.* The magnetization and $H_k$ were detected with the anomalous Hall effect (AHE). All the Hall resistances ($R_H$) were detected by applying a 50 μA dc current. For the SOT switching, a 1 ms current pulse with the desired amplitude was applied to switch the magnetization, and then $R_H$ was recorded. All electrical measurements were performed after the samples cooled down to room temperature if heated.

*STEM measurements.* Cross-section STEM samples were prepared using a Zeiss Auriga focused ion beam (FIB) system. STEM imaging was performed on an FEI Titan with CEOS probe aberration corrector operated at 200 kV with a probe convergence angle of 24.5 mrad, spatial resolution of 0.08 nm, and probe current of ~20 pA. The STEM samples were loaded into the microscope immediately after the FIB sample preparation to avoid possible oxidization of the Co in air.

## III. Results and discussion

### A. STEM observation of voltage-driven oxygen migration in metallic systems

To directly observe EF-modulated metallic HM/FM interface through ionic effects, we performed STEM and electron energy loss spectroscopy (EELS) measurements in a 4 nm Co layer with the structure of Si/SiO$_2$ (300 nm)/Pt (10 nm)/Co (4 nm)/GdO$_x$ (50 nm). As shown in Fig. 1(a), the effective voltage control area was patterned into a 200 × 200 μm$^2$ square. Figure 1(b) shows $R_H$ as a function of the applied vertical magnetic field ($H_z$) at different voltage control states. The as-deposited state shows an in-plane anisotropy state (sample c in Fig. 1(b)), which could be modified to a PMA state (sample d and f) or a fully oxidized state (sample e) by applying a gate voltage ($V_G$) [15]. For sample d and e, only $V_G$ = -5 V was applied. For sample f, a $V_G$ = -5 V was applied first until a fully oxidized state was achieved, and then a $V_G$ = +5 V was applied to rebuild PMA as shown in previous works [15]. Figure 1(c-f) show the depth distribution of oxygen and Co near the Pt/Co interface. For the as-deposited state (Fig.

1(c)), oxygen ions ($O^{2-}$) only distribute near the Co/GdO$_x$ interface and the Pt/Co interface is clean (no $O^{2-}$ detected). After $V_G$ = -5 V application, the Co and oxygen curves have overlapped each other at the Pt/Co interface (Fig. 1(d)), but the electrical measurement still shows a clear $R_H$ signal from Co metal at this state (Fig. 1(b)), indicating that the Co is only partially oxidized and $O^{2-}$ can penetrate the unoxidized Co metal and reach the Pt/Co interface. Therefore, the combination of both STEM and electrical measurement results can directly confirm that the EF-driven $O^{2-}$ can diffuse into the Co metal layer. For a fully oxidized state, the Co and oxygen curves (Fig. 1(e)) are similar with Fig. 1(d), but there is no $R_H$ signal detected (Fig. 1(b)), indicating a fully oxidized Co layer.

Figure 1(f) shows the Co and oxygen distributions for sample f. As mentioned above, this sample was fully oxidized first by applying $V_G$ = -5 V and then was reduced to a PMA state by applying $V_G$ = +5 V. Although both sample d and f show a PMA state with the same $R_H$, the oxygen signal sharply drops to almost zero before reaching the Pt/Co interface in sample f, in contrast to sample d in which the oxygen signal drops together with the Co signal at the Pt/Co interface. By comparing the oxygen distributions in these two samples, we can conclude that sample f has a much cleaner Pt/Co interface than sample d. Since the Pt/Co interface is only manipulated by applied $V_G$ in both samples, these results indicate that the metallic Pt/Co interface can be controlled through EF-driven $O^{2-}$ motion. A slight oxygen signal at $z \approx 2$ nm in Fig. 1(f) may be due to a small amount of trapped oxygen [25] that cannot be removed under -$V_G$.

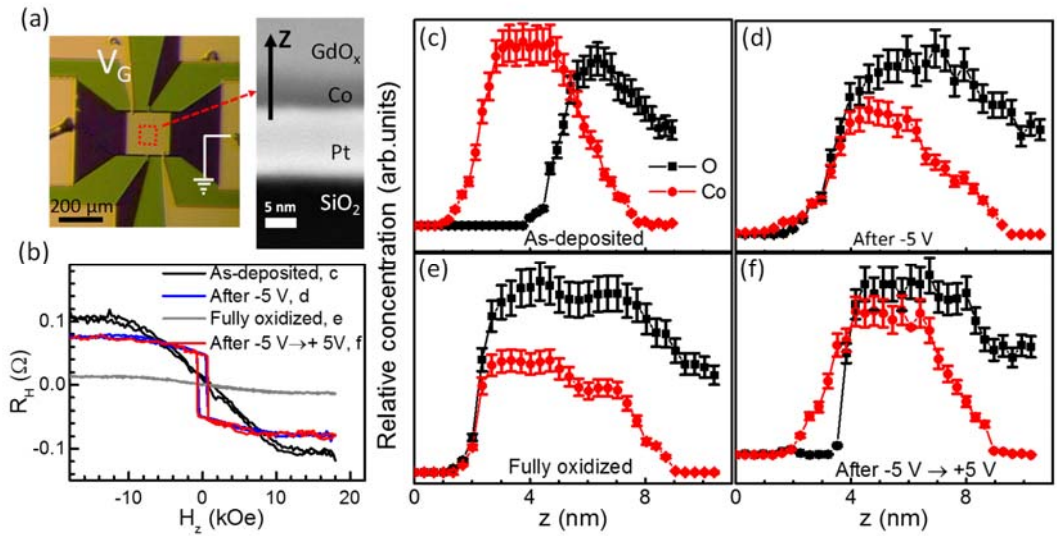

FIG. 1. STEM EELS results at different EF-controlled states. (a) Top-view and typical STEM image of a Pt/Co/GdO$_x$ sample. (b) The vertical field dependent $R_H$ for samples c, as-deposited state, d, after $V_G$ = -5 V application, e, fully oxidized state, and f, after $V_G$ = -5 V and then $V_G$ = +5 V applications. (d-f) The corresponding Co and oxygen distribution from EELS line scan along z direction at the Pt/Co interface.

**B. Schematic of voltage-controlled magnetic states at HM/FM interfaces**

Voltage control of metallic HM/FM interfaces must be also accompanied by the control of FM/oxide interfaces in HM/FM/oxide structures, and moreover, the alteration of FM/oxide interfaces may dominate the overall control effects. Therefore, determining the contribution from HM/FM interfaces during the control process becomes a critical step in investigating the voltage control of HM/FM interfaces. A schematic of the voltage control of metallic HM/FM interfaces is presented in Fig. 2. Here we assume that the PMA originates from both the Co/GdO$_x$ interface [4,32,33] and Pt/Co interface [2,6–8,17] in Pt/Co/GdO$_x$ structures. As observed in Fig. 1, O$^{2-}$ can penetrate the metallic Co layer and reach the Pt/Co interface before the Co is fully oxidized. Figure 2(a-e) schematically show the voltage gated O$^{2-}$ migration by considering O$^{2-}$ diffusion. The expected perpendicular component of magnetization (M$_z$) under both in-plane and vertical fields are shown in Fig. 2(f-j), respectively. We classify the O$^{2-}$ in this system into two different types, depending on the strength of the bonding energy between O$^{2-}$ and metal atoms. Type-I is the O$^{2-}$ with a weak bond to metal atoms and can be moved along grain boundaries through self-diffusion or be driven by an electric field [25,33–36]. Type-II is the O$^{2-}$ strongly bonded to metal atoms, and the significant migration of this type of O$^{2-}$ may require a strong electric field and an elevated temperature [15,37].

As shown in Fig. 2(a), in the initial state, the Co layer is partially oxidized but the Pt/Co interface is clean. This state usually shows a strong H$_k$ because of the orbital hybridization between Co and type-II O$^{2-}$ at the top surface of the Co layer as well as the clean Pt/Co interface. When applying a +V$_G$, the O$^{2-}$ is driven out of the Co/CoO$_x$ or Co/GdO$_x$ interface as shown in Fig. 2(b), resulting in a decreasing H$_k$ and an increasing M$_s$ through the reduction of Co from CoO$_x$ [15]. With increasing +V$_G$ duration, all the O$^{2-}$ can be depleted at the Co/GdO$_x$ interface and the entire Co layer becomes fully reduced to an in-plane anisotropic pure Co metal [15,37] (Fig. 2(c)). Then, when -V$_G$ is applied, the O$^{2-}$ is driven back to the Co/GdO$_x$ interface (Fig. 2(d)). Type-II O$^{2-}$ is formed at the Co/GdO$_x$ interface and PMA is recovered. However, by considering the diffusion of type-I O$^{2-}$ through grain boundaries, the situations of Fig. 2(d) and Fig. 2(b) will be different. It is expected that type-I O$^{2-}$ can penetrate the Co layer, and thus the Pt/Co interface in Fig. 2(d) will be contaminated by type-I O$^{2-}$. By considering oxygen contamination at the Pt/Co interface and the contribution of the Pt/Co interface to the PMA, the PMA of Fig. 2(d) will be much weaker than that in Fig. 2(b) with a clean Pt/Co interface, even though the other conditions are exactly the same. Figure 1(d) and 1(f) have directly confirmed the O$^{2-}$ distribution in the two dramatically different situations, and together with magnetic measurements below, these results will provide clear evidence of electrical control of metallic Pt/Co interfaces that has not been explored before. The Co layer can be fully oxidized for a longer -V$_G$ application time (Fig. 2(e)). If +V$_G$ is then applied again, type-I O$^{2-}$ at the Pt/CoO$_x$ interface will be moved first, and then the CoO$_x$ layer will be gradually reduced. Since type-I O$^{2-}$ has been removed near the Pt/CoO$_x$ interface before the reduction of Co, the sample returns to the state of Fig. 2(b). Because M$_s$ mainly depends on the oxidization/reduction of the Co layer at the Co/GdO$_x$ interface while H$_k$ is determined by both the Pt/Co and Co/GdO$_x$ interfaces, M$_s$ and H$_k$ can be controlled independently through V$_G$ modulated O$^{2-}$ distribution at the two interfaces.

In experiments, we show that all these states in Fig. 2(a-e) can be reliably controlled in the sequence of a → b → c → d → e → b, and the magnetism of the Co can be manipulated to any possible state. Moreover, two regions with a constant M$_s$ around the pure metallic Co state (Fig. 2(c)) and a constant $H_k$ during the $-V_G$ application (Fig. 2(d)) are obtained. These two special regions are further used for investigating SOT switching by altering either $M_s$ or $H_k$ independently.

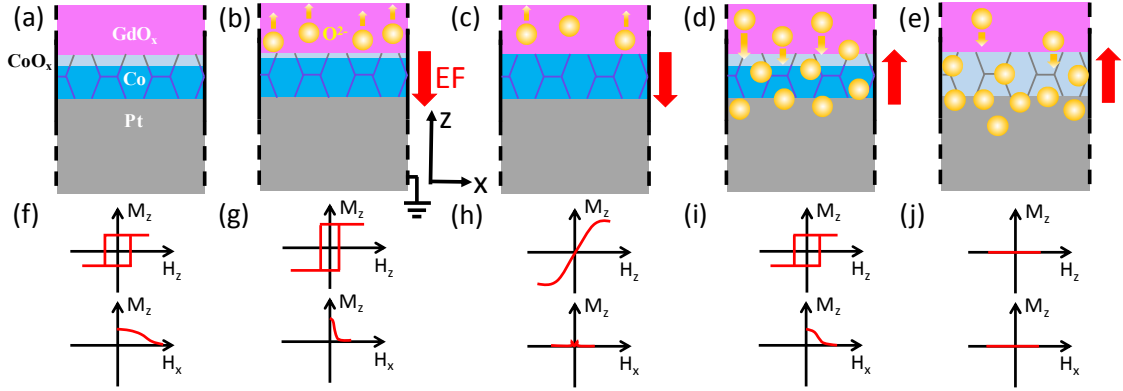

FIG. 2. Schematic of the voltage control of magnetism through metallic Pt/Co interfaces. (a) The as-deposited state showing PMA without $V_G$ application. The Co layer has been partially oxidized and the Pt/Co interface is clean. (b) Under $+V_G$ application, the $O^{2-}$ at the Co/CoO$_x$ or Co/GdO$_x$ interface is driven into GdO$_x$ layer. The CoO$_x$ is partially reduced to Co and the Pt/Co interface keeps clean. (c) The CoO$_x$ is finally reduced to Co and the Co layer becomes in-plane anisotropic under the $+V_G$ application. (d) After $-V_G$ application, the $O^{2-}$ is driven back to the Co/GdO$_x$ interface and further contaminates the Pt/Co interface through diffusion. PMA is rebuilt and the Co layer is oxidized. Due to the contamination of the Pt/Co interface, $H_k$ becomes weaker compared to (b) even with the same other conditions. (e) The Co layer is fully oxidized by $-V_G$ application. The blue line texture in the Co/CoO$_x$ layer symbolically illustrates grain boundary. (f-j) The schematic of expected vertical and in-plane field dependent $M_z$ at each state.

## C. Experimental demonstration of voltage-controlled metallic Pt/Co interfaces

The samples for demonstrating the voltage control effects are the Si/SiO$_2$ (300 nm)/Pt (3 nm)/Co (0.9 nm)/GdO$_x$ (80 nm) samples, which were patterned into Hall bar structures with a feature size of 2.5 μm. A schematic of the sample structure is shown as the inset of Fig. 3(d). $V_G$ was applied at an elevated temperature between 200 - 260 °C while the magnetic states were detected at room temperature. Figure 3(a) and (b) show the typical vertical and in-plane field dependent $R_H$, respectively. In Fig. 3(b), the $H_x$ was tilted about 4° toward the $z$ direction to get a coherent magnetization switching. The $M_s$ which is proportional to $R_H$ [38] can be obtained from Fig. 3(a) and $H_k$ can be evaluated from Fig. 3(b) through the Stoner-Wohlfarth model [39]. We first confirmed that the magnetism could be controlled by $V_G$ and then extracted $R_H$ and $H_k$ as a function of $V_G$ application time as shown in Fig. 3(c), where $V_G$ was applied at 260 °C. As shown in Fig. 3(c), from the as-deposited state ($R_H$ = 0.15 Ω; $H_k$ = 8.95 kOe), the sample was first set to a fully oxidized state with almost zero $R_H$ and $H_k$ under $-V_G$ application (first violet area on the left). Then, under $+V_G$ application, $R_H$ monotonically increases to a maximum value of 0.48 Ω while $H_k$ first increases and then decreases. The monotonical increase of $R_H$ indicates that the CoO$_x$ is continuously reduced to Co. The increase of $H_k$ at the initial stage illustrates that the PMA is gradually established as $O^{2-}$ is removed from the CoO film, and the subsequent decrease of $H_k$ is due to a thicker Co layer that prefers in-plane anisotropy [15]. Next, under $-V_G$ application, the sample returns to the PMA state and then to a fully oxidized state. The magnetic transition between the pure metallic state and the fully oxidized state can be reliably repeated, and all of these results are consistent with previous reports [15]. Note the magnetic states near the fully oxidized state, with zero $R_H$ and $H_k$ after $-V_G$ application, which were attributed to in-plane anisotropic states [16] probably correspond to superparamagnetic states [39].

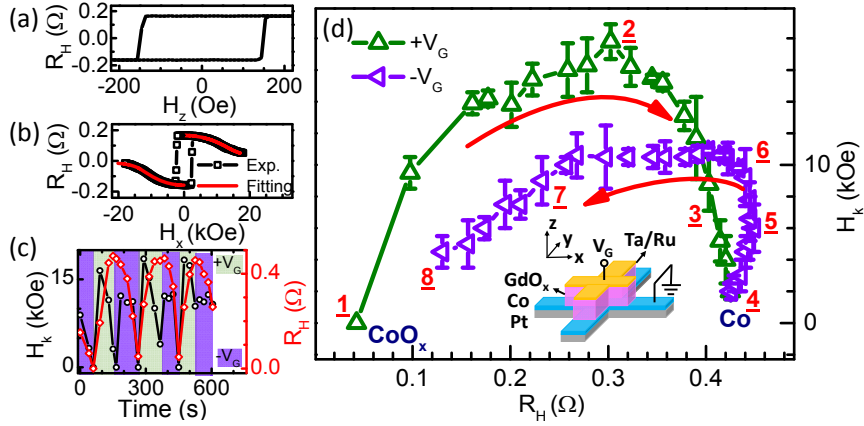

FIG. 3. The evolution of magnetism under $V_G$ control. (a, b) Typical $R_H$ curves under vertical (a) or in-plane fields (b). The red lines in (b) are fitting results by using the Stoner-Wohlfarth model. (c) Typical time dependences of $H_k$ and $R_H$ under $V_G$ control. (d) The measured $H_k$ as a function of $R_H$ in a typical $V_G$ control cycle. The red arrows indicate the time evolution of the magnetism from state 1 to 8. Inset: the sample geometry and the experimental configuration of $V_G$ application.

We then inspected the detailed evolution of $H_k$ and $M_s$ with $V_G$ applied at 200 °C. The lower temperature allows a finer control of $H_k$ and $M_s$ by voltage. In Fig. 3(d), we plot the extracted $H_k$ as a function of $R_H$ at each state after $V_G$ application. As indicated by the red arrows, the $V_G$ control cycle starts from an almost fully oxidized state (state 1) goes to a pure metallic state (state 4) under $+V_G$ application, and then back to another almost fully oxidized state (state 8) under $-V_G$ application. The remarkable feature of Fig. 3(d) is that the $H_k$-$R_H$ curve shows a hysteresis-like behavior, where $H_k$ has different values after $+V_G$ and $-V_G$ applications at the same $R_H$. Particularly, two regions are distinctive: from state 4 to 6, $H_k$ sharply increases from 0 to about 10 kOe while $R_H$ keeps almost the same; from state 6 to 7, $H_k$ keeps almost constant over a broad range of $R_H$.

The hysteresis-like $H_k$-$R_H$ curve is significant because the same $R_H$, which corresponds to the same oxidization level of the Co layer and thus the same Co/oxide interface, should have the same $H_k$, if only the PMA at the Co/oxide interface is controlled by voltage. To explain these results, the Pt/Co interface must also be modified by the voltage through $O^{2-}$ diffusion. According to Fig. 2, the states between 1–2 in Fig. 3(d) have a clean Pt/Co interface after $+V_G$ application, corresponding to Fig. 2(b) with a larger $H_k$, and the states between 7–8 have a contaminated Pt/Co interface after $-V_G$ application, corresponding to Fig. 2(d) with a weaker $H_k$. Different $H_k$ values at the same $R_H$ can be well understood in this way. This explanation is also directly confirmed by the STEM samples. The measured $H_k$ of sample f in Fig. 1 is 4.3 kOe, larger than that of sample d ($H_k$= 2.1 kOe), even with the same $R_H$ as shown in Fig. 1(b). Compared to sample d, sample f has a cleaner Pt/Co interface as shown in Fig. 1(d) and 1(f), well consistent with this explanation that the origin of larger $H_k$ at the same $R_H$ arises from a cleaner Pt/Co interface. Nevertheless, the $H_k$ behaviors in the right minor loop $\underline{3}$-$\underline{4}$-$\underline{5}$-$\underline{6}$ cannot be simply explained through Fig. 2(b) and Fig. 2(d). In this minor loop, $H_K$ first quickly increases from $\underline{4}$ to $\underline{6}$ after -$V_G$ application and then reaches a saturation value while $R_H$ keeps almost constant. After that, $M_s$ decreases while $H_k$ maintains nearly constant ($\underline{6}$ to $\underline{7}$). Beyond $\underline{7}$, both $M_s$ and $H_k$ decrease when the Co film is gradually returned to CoO by -$V_G$. Contrary to the left major loop, $H_k$ after –$V_G$ application in the minor loop is larger than that after +$V_G$ application at the same $R_H$.

To explain the minor loop, we consider the magnetic states starting from the metallic Co state $\underline{4}$ in which all the $O^{2-}$ has been driven into the $GdO_x$. In the beginning of -$V_G$ application ($\underline{4}$ to $\underline{5}$), the -$V_G$ driven $O^{2-}$ moves toward the Co/$GdO_x$ interface and bonds to the Co atoms to build PMA. Since the $O^{2-}$ concentration is still very low at this initial state, the Co layer has not been significantly oxidized, consistent with the fact that $R_H$ keeps almost constant and $H_k$ increases as shown in Fig. 3(d). Additionally, the $O^{2-}$ has not diffused into the Co layer and the Pt/Co interface keeps clean. Compared to the states between $\underline{3}$-$\underline{4}$, the larger $H_k$ between $\underline{4}$-$\underline{5}$ are likely due to these $O^{2-}$ at the Co/$GdO_x$ interface that contribute to the PMA but do not significantly oxidize the Co layer. For the states between $\underline{3}$-$\underline{4}$, at the final stage of +$V_G$ application, all the $O^{2-}$ is depleted except those with strong a bond to Co in unreduced $CoO_x$, and thus there is no such $O^{2-}$ only contributing to the PMA (all the $O^{2-}$ strongly bonds to Co and contributes to the $CoO_x$ for the states between $\underline{3}$-$\underline{4}$) and $H_k$ is smaller. From state $\underline{5}$ to $\underline{6}$, the increased rate of $H_k$ change starts to decrease although the $H_k$ has not been saturated, and $R_H$ begins to decrease. These magnetic states indicate that the $O^{2-}$ has diffused to the Pt/Co interface and concentrated enough to start oxidizing the Co layer. The appearance of the minor loop further confirms the $O^{2-}$ diffusion and the contribution from both the Pt/Co and Co/$GdO_x$ interfaces to the voltage control of PMA.

In principle, by controlling the $O^{2-}$ at the both the Pt/Co and Co/$GdO_x$ interfaces, all the possible magnetic states shown in the major and minor loops of Fig. 3(d) can be achieved. Figure 4(a) shows the controlled magnetic states starting from the same initial magnetic state with zero $R_H$ and $H_k$, but the -$V_G$ are applied from different states. The controlled magnetic states where -$V_G$ are applied from $R_H$ = 0.33 Ω, 0.37 Ω, and 0.41 Ω are plotted in purple, red, and violet curves, respectively. The gray curve is from Fig. 3(d) for comparison. This figure shows that the states within the major and minor loops can be achieved by controlling the beginning of the -$V_G$ application. Figure 4(b) presents the magnetic states under +$V_G$ application but starting from different initial states. These initial states are obtained through –$V_G$ applications. Thus, this plot presents the control effects that the beginning of +$V_G$ application leads to. It also shows that the controlled magnetic states depend on the beginning state where the reversal $V_G$ is applied. Moreover, Fig. 4(b) shows a larger initial $R_H$ value can result in a larger maximum $H_k$ in the following control process. These results indicate that the magnetism can be controlled to any possible magnetic state by controlling the $V_G$ application process.

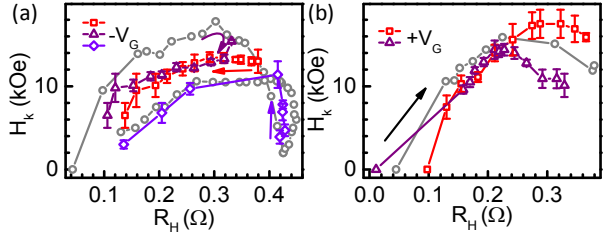

FIG. 4. The effects of reversal $V_G$ applications on the control effects. (a) $H_k$ versus $R_H$ curves when $-V_G$ is applied from $R_H = 0.33$ Ω (purple), 0.37 Ω (red), and 0.41 Ω (violet). (b) $H_k$ versus $R_H$ curves where $+V_G$ is applied from different initial magnetic states. The gray curve is from Fig. 3(d) for comparison. The arrows in (a, b) indicate the time evolution of controlled magnetic states.

### D. Voltage control of SOT switching

SOTs offer a very promising way to switch magnetization electrically [11,40]. However, many aspects of SOT switching are still under debate. For example, the origin of SOTs can be spin Hall effects (SHEs) [40–42] or Rashba effects [11,43], and their roles during the switching process are still not clear [11,41,42,44–46]. Generally, the SOT switching starts from a reversal domain nucleation in a uniform magnetization state and then the nucleated domains expand to the entire film [44,47,48]. Therefore, the critical SOT switching current corresponds to the current for destabilizing the uniform magnetization, which is attributed to damping-like [42,49] or field-like spin torques [45] as well as additional thermal effects [50–52]. Extracting the intrinsic magnetic property dependence of the SOT switching in the same sample will provide critical insights on the switching mechanism. However, it cannot be realized through a conventional method since the intrinsic properties are specified during the sample fabrication. As demonstrated above, the $M_s$, $H_k$, as well as the oxidation level at both interfaces of the Pt/Co/GdO$_x$ structures can be independently controlled by $V_G$, thereby offering a unique approach for studying SOT effects.

The voltage controlled SOT switching was explored in samples with the structure of Si/SiO$_2$ (300 nm)/Pt (3 nm)/Co (0.6 nm)/GdO$_x$ (80 nm) multilayers. The SOT switching was measured in the same way as previous studies [44,46,51]. Figure 5(a) shows typical current driven SOT switching curves under in-plane fields. The magnetization switching direction depends on both the current and field directions, which is consistent with previous reports [11,40,41,44,46–48,51]. We choose the current where $R_H$ begins to change as $I_c$, as marked in Fig. 5(a). As mentioned above, $I_c$ corresponds to the critical current for destabilizing a uniform magnetization, under which the reversal domains start to nucleate. Figure 5(b) shows the switching curves after -$V_G$ and then +$V_G$ applications. One can see that $I_c$ can be greatly decreased by -$V_G$ together with the reduced $M_s$, and increased by +$V_G$ in the opposite fashion.

Figure 5(c) shows the evolution of $H_k$ and $R_H$ under the ±$V_G$ control. The $H_k$-$R_H$ curve exhibits a similar hysteretic behavior as in Fig. 3(d). During the application of -$V_G$, $H_k$ can be maintained nearly constant at 5 kOe over a large $R_H$ region (0.15-0.35 Ω, states between 6-7). A minor loop also exists near the pure Co state, where a large change of $H_k$ (from 2 kOe to 5 kOe) is accompanied by a small change of $R_H$ (states between 4-5). $I_c$ is plotted against $R_H$ in Fig 5d. Remarkably, $I_c$ can be reduced by more than 10 times with the application of -$V_G$, demonstrating the advantage of controlling SOT switching by ionic effects. The modulation of $I_c$ is much larger than that by using an ionic liquid gate [53] in which only the FM/oxide interface was controlled. Particularly, in the two special regions, it is clearly shown that $I_c$ is proportional to $R_H$ (states 6-7) and also strongly depends on $H_k$ (states 4-5).

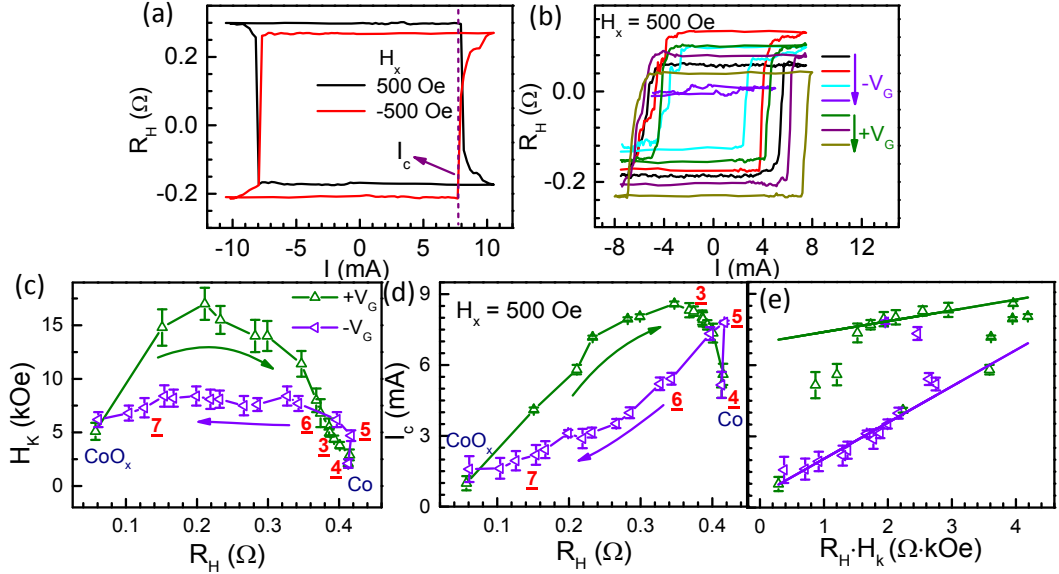

FIG. 5. SOT switching under $V_G$ control. (a) Typical current induced SOT switching under in-plane fields. (b) The SOT switching curves under $V_G$ control. (c, d) The extracted $H_k$ (c) and corresponding $I_c$ (d) as a function of $R_H$ at each magnetic state controlled by $V_G$. The arrows indicate the time evolution of control process. (e) $I_c$ is replotted as a function of $R_H \cdot H_k$. The solid lines are the liner fitting results.

To analyze the SOTs and their effects on the magnetization switching, we adopt a simple magnetic switching model by only considering damping-like torque [42,50]. Although the field-like torques [45] and possible DMI [54] are not counted, this simple model can explain several experimental observations [55,56] and thus may provide some initial insights into our results. According to this model [42,50],

$$I_c = \frac{2e}{\hbar} \frac{M_s t_F}{\theta_{SH}} (\frac{H_k}{2} - \frac{H_x}{\sqrt{2}}), \quad (1)$$

where $\theta_{SH}$ is the effective spin Hall angle and $t_F$ is the thickness of ferromagnetic layer. By considering $R_H \propto M_s t_F$ and $H_x \ll H_k$,

$$I_c \propto \frac{R_H H_k}{\theta_{SH}}, \quad (2)$$

where the slope of $I_c$ versus $R_H \cdot H_k$ represents the $\theta_{SH}$ in this system. Figure 5(e) shows the measured $I_c$ as a function of $R_H \cdot H_k$ by replotting Fig. 5(c) and (d), and the solid lines are the linear fitting results. Note the data points obtained after the application of +$V_G$ and –$V_G$ are plotted as up triangles and left triangles, respectively. Immediately two linear regions can be identified, generally confirming the validity of this simple model. The first linear region is in the range of $I_c < 5.5$ mA, in which the slopes of $I_c$ versus $R_H \cdot H_k$ is about $1.57 \times 10^7$ A cm$^{-2}\Omega^{-1}$kOe$^{-1}$. Another linear region is within the range of 7.0 mA $< I_c <$ 9.0 mA. This region corresponds to the states with a clean Pt/Co interface after $+V_G$ application. The slope in this region is about $0.47 \times 10^7$ A cm$^{-2}\Omega^{-1}$kOe$^{-1}$. This slope corresponds to a 3.5 times larger $\theta_{SH}$ than that after $-V_G$ application, demonstrating that a clean Pt/Co interface plays a critical role in the efficiency of SOTs. Note the Joule heating induced thermal effects during the magnetization switching cannot explain the larger $\theta_{SH}$ after $+V_G$ application [39].

## IV. Conclusions

HM/FM interfaces play a critical role in many interfacial phenomena, such as spin Seebeck effects [29], spin Hall magnetoresistance [30], and other magnetic proximity effects [31]. We have demonstrated that the metallic HM/FM interface can also be controlled electrically, providing new opportunities for electrical modulation of these interfacial effects. Contrary to previous studies where only the FM/oxide interface was controlled and the effect was limited to a few special states, the controlled HM/FM interface strongly broadens the magnetic states in the field of voltage controlled magnetism. In addition, $M_s$ and $H_k$ can be manipulated independently and the magnetism can be set to any possible magnetic state. STEM and EELS results directly confirm the controllable metallic interface and its effects on the control of $H_k$ at the same $R_H$.

Recently, it has been shown that the PMA in a HM/FM/metal/Oxide structure strongly depends on the thickness of the top capping metal layer [16,36,57]. The speculated reasons focused only on the influence of the capping metal layer on the FM/metal interface. This work indicates an alternative mechanism that the thickness of the capping layer may alter the $O^{-2}$ distribution underneath at both the HM/FM and FM/metal interface during the sputtering process. By considering the contributions to the PMA from both interfaces, the change of the $O^{-2}$ distribution at the two interfaces may result in the capping metal layer dependence on the PMA.

Furthermore, we show that the efficiency of SOTs can also be controlled and the critical SOT switching current can be reduced about 10 times. Previous studies have indicated that the efficiency of spin injection at the interface depends on the transparency of this interface [27,28], but demonstrated in different samples that inevitably also alter other conditions. Our results clearly show that the effective SOTs can be enhanced about 3.5 times with a clear HM/FM interface in the same sample. The unique voltage control of magnetism may also be applied to investigate other magnetism-related effects by modulating the $M_s$ and $H_k$ independently in the same sample, avoiding the influence of the variation between different samples in conventional measurements.

## Acknowledgements

This work was supported by C-SPIN, one of six centres of STARnet; a Semiconductor Research Corporation programme, sponsored by MARCO and DARPA; and by the National Science Foundation through ECCS-1554011.

**Supplemental Material for**

Electrical control of metallic heavy-metal/ferromagnet interfacial states

Chong Bi[1*], Congli Sun[2], Meng Xu[1], Ty Newhouse-Illige[1], Paul M. Voyles[2], and Weigang Wang[1†]

[1]Department of Physics, University of Arizona, Tucson, Arizona 85721, USA.

[2]Department of Materials Science and Engineering, University of Wisconsin-Madison, Madison, WI 53706, USA

[*]cbi@email.arizona.edu

[†]wgwang@physics.arizona.edu

**1. Extracting $H_k$ through the Stoner-Wohlfarth model**

**2. Anisotropic magnetoresistance at each magnetism state**

**3. Thermal effects during SOT switching**

## 1. Extracting $H_k$ through the Stoner-Wohlfarth model

The stoner-Wohlfarth model is widely used to describe the rotation of a uniform magnetization under an external field. To evaluate $H_k$, the coherent rotation of magnetization under varied in-plane fields needs to be kept, especially for the magnetism states with very weak PMA. To do this, we tilt the in-plane field about 4° toward the normal direction of the film, which is enough for keeping coherent switching for most of magnetism states after $V_G$ control. We have also measured $H_k$ at other tilt angles from 1° to 10° and found the $H_k$ error is within 5% for the different tilt angles.

We consider a simple ferromagnetic system, in which the total energy includes the perpendicular anisotropy energy $E_A$ and the Zeeman energy $E_z$. The total energy E at 0 K can be written as

$$E = E_A + E_Z = K\cos^2\theta - HM_s\cos(\theta - \varphi) \qquad (1)$$

where $K$ is the anisotropy constant, $\theta$ is the angle between the magnetization and film plane, $H$ is the external field, $M_s$ is the saturation magnetization, and $\varphi$ is the tilt angle of the external field relative to the film plane. Minimization of the energy ($\delta E/\delta\theta=0$) yields the relation

$$2K\cos\theta - HM_s(\cos\varphi - \sin\varphi\cot\theta) = 0. \qquad (2)$$

According to Eq. 2, we can extract $H_k$ by fitting experimental data as shown in Fig. 2(b).

## 2. Anisotropic magnetoresistance at each magnetism state

In addition to $R_H$ measurements, we also measure the anisotropic magnetoresistance (AMR) by using a Hall bar structure with the size of 10 μm × 50 μm at different magnetic states. Figure S1(d-f) show the AMR results at three typical magnetic states: as-deposited state, after $-V_G$ application until $R_H$ versus $H_z$ shows paramagnetic hysteresis-like behaviors, and after $+V_G$ application also until $R_H$ versus $H_z$ shows paramagnetic hysteresis-like behaviors. Figure S1(a-c) show the corresponding $R_H$ versus $H_z$ curves of (d-f), respectively. It should be noted that the $R_H$ of Fig. S1(b) is much smaller than that of Fig. S1(c). The clear AMR signals of the magnetism state after $+V_G$ application indicate an in-plane anisotropy at this state. No AMR signal is observed for the state after $-V_G$ application indicating that the magnetic state is probably in a superparamagnetic state.

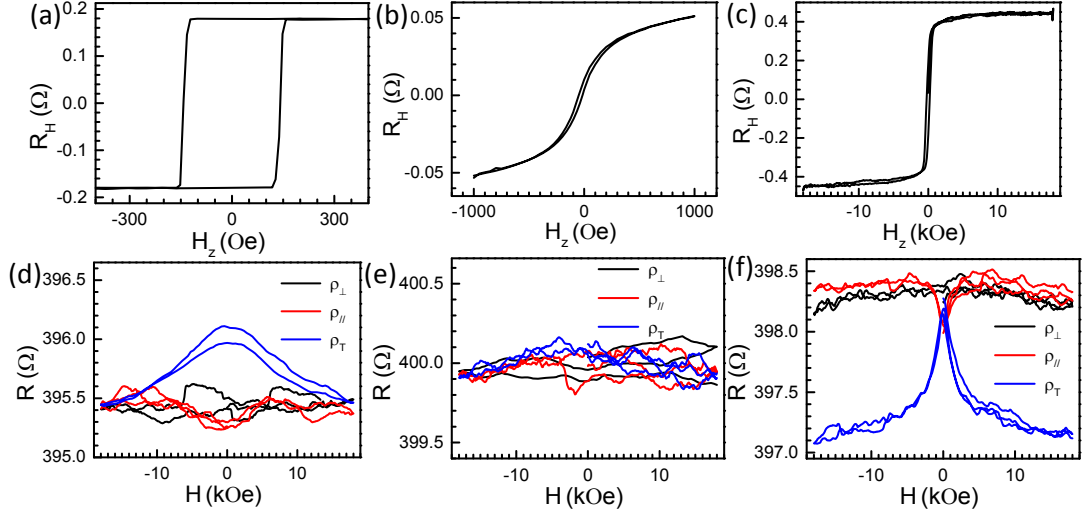

FIG. S1. AMR measurement results after $V_G$ control. (a-c) $R_H$ versus $H_z$ for as-deposited state (a), after – $V_G$ application (b), and after +$V_G$ application (c). (d-e) The corresponding AMR results of (a-c), respectively. The external field is applied perpendicular to the film plane $\rho_\perp$, in-plane parallel with current direction $\rho_{//}$, and in-plane perpendicular with current direction $\rho_T$.

## 3. Thermal effects during SOT switching

The duration of the applied current pulses for SOT switching is about 1 ms, which is long enough for detecting thermal effects by using an oscilloscope. Since the main material of the sample is Pt, which shows a stable and high temperature coefficient, the temperature of the sample during pulse durations can be monitored by the resistance of the sample. Before measurements, the temperature coefficient of the sample, which is about $1.03 \times 10^{-3}$ K$^{-1}$, is calibrated. Figure S2(b) shows the time dependent voltage acquired by an oscilloscope during a 1 mA current pulse. According to the voltage amplitude ($V_{Mea}$), we can calculate the resistance of the sample during the current pulse. Figure S2(c) shows the calculated resistance as a function of pulse current and the square of pulse current (inset). A liner relation between R and $I^2$ indicates that the increase of R is mainly due to Joule heating. The temperature increase ($\Delta T$) as a function of applied pulse current is shown in Fig. S2(d), which shows the temperature increase as high as 191 K during a 10 mA current pulse. These thermal effects can assist in the SOT switching [50, 51], however, they cannot explain the smaller slope observed in the range of 7 mA < $I_c$ < 9 mA (Fig. 5(e)). In the range of 0 mA < $I_c$ < 5.5 mA, $\Delta T$ is about 50 K, and in the range of 7 mA < $I_c$ < 9 mA, $\Delta T$ is about 60 K. Because both regions have almost the same temperature increase, the contributions of the thermal effects to the change of slope should be almost the same [50].

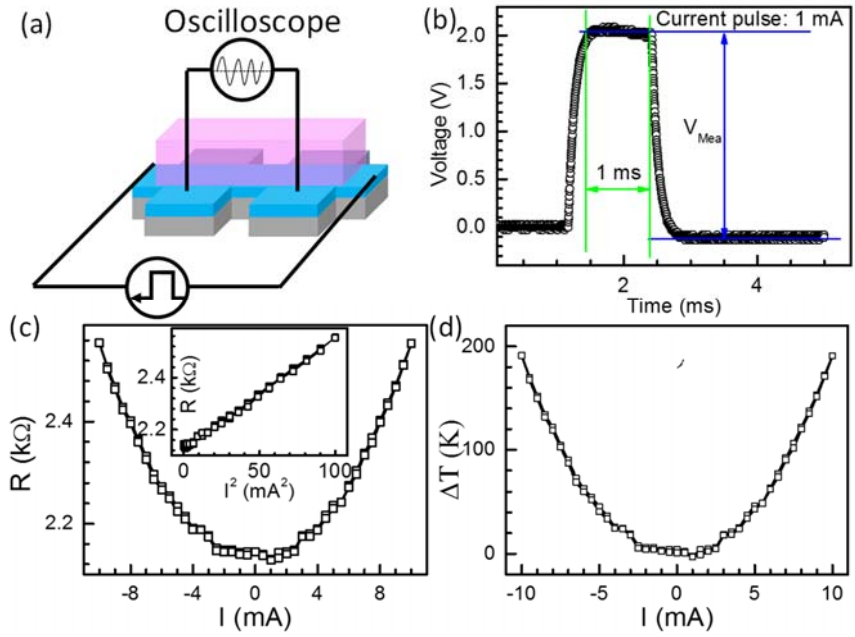

FIG. S2. Estimation of the temperature increase during applied current pulses. (a) Schematic of the experimental configuration for the time dependence measurements. (b) The measured time dependent voltage when the current pulse is 1 mA. (c) The resistance of the sample as a function of pulse current. The inset shows a linear relation of R versus $I^2$. (d) The calculated temperature increase as a function of applied current.